\documentclass[12pt,preprint]{aastex}

\shorttitle{GRBs: Restarting the Engine}
\shortauthors{Andrew King et al.}

\begin{document}
\def\grb{{GRB050502b}}
\def\lsun{{\rm L_{\odot}}}
\def\msun{{\rm M_{\odot}}}
\def\rsun{{\rm R_{\odot}}}
\def\go{
\mathrel{\raise.3ex\hbox{$>$}\mkern-14mu\lower0.6ex\hbox{$\sim$}}
}
\def\lo{
\mathrel{\raise.3ex\hbox{$<$}\mkern-14mu\lower0.6ex\hbox{$\sim$}}
}
\def\simeq{
\mathrel{\raise.3ex\hbox{$\sim$}\mkern-14mu\lower0.4ex\hbox{$-$}}
}

\title{Gamma-ray bursts: Restarting the Engine}
\author{Andrew King, Paul T. O'Brien, Michael R. Goad, Julian Osborne,\\
Emma Olsson, Kim Page}
\affil{Department of Physics \& Astronomy,
University of Leicester, Leicester LE1 7RH, UK}

\begin{abstract}
Recent gamma--ray burst observations have revealed late--time, highly
energetic events which deviate from the simplest expectations of the
standard fireball picture. Instead they may indicate that the central
engine is active or restarted at late times. We suggest that
fragmentation and subsequent accretion during the collapse of a
rapidly rotating stellar core offers a natural mechanism for this.
\end{abstract}

\section{Introduction}

It is now widely believed that gamma-ray bursts (GRBs) lasting longer than
about 1~s are produced by a class of supernovae, called hypernovae
(Paczy\'nski, 1998; MacFadyen \& Woosley, 1999). Collapse of a massive,
rapidly-rotating stellar core is assumed to lead to the formation of a black
hole, while the remaining core material has enough angular momentum to form a
massive accreting neutron torus around it. The infalling torus radiates its
gravitational energy as neutrinos or converts it directly to a beam by MHD
processes. This evacuates the rotation axis of the core, allowing both an
observable burst of gamma rays and the expulsion of a jet of matter at high
Lorentz factors. The interaction of this jet with the surroundings produces
the X-ray afterglow, which has proved so useful in locating and studying
gamma-ray bursts.

In all cases observed until now, once the initial gamma-ray emission
has faded away the energy output is dominated by the afterglow which
has far lower total energy than the GRB.  However, the recent \grb \
is quite different (Burrows et al. 2005). An X-ray flare occurred
starting some 400 s after the beginning of the GRB, and released at
least as much energy as the original burst. There have been several
efforts to explain \grb \ by adding elements to the standard picture
of long GRBs such as late internal shocks, or synchrotron
self--Compton emission (Burrows et al.  2005; Kobayashi et
al. 2005). However there is considerable difficulty in understanding
how \grb \ can put so much energy into its surroundings at such late
times, and what physical parameter specifies this very different
behaviour.

Here we consider an idea which offers a way out of these
difficulties. In effect it offers a physical model for the late
internal shock suggested by Burrows et al. (2005).

\section{Observations}

GRB050502b was detected by the Burst Alert Telescope (BAT) on the {\it
Swift} satellite (Gehrels et al. 2004) at 09:25:40 UT on 2005 May 2
(Falcone et al. 2005). The burst lasted 17.5 s (T90) with a fluence in
the 15--350 keV band of $(8\pm 1) \times 10^{-7}$ erg cm$^{-2}$
(Cummings et al. 2005). The gamma-ray spectrum is well fitted by a
single power law with a photon index, $\Gamma = 1.6\pm 0.1$. Falcone
et al. reported that initially the X-ray count rate seen by {\it
Swift} was too low to obtain an on-board centroid position with the
{\it Swift} X-ray Telescope (XRT) in the 0.3--10 keV band, but
subsequent ground-processing revealed an initially fading X-ray
source. This X-ray source brightened dramatically starting some 400 s
after the initial burst, rising to a peak which lasted for several
hundred seconds before fading away (Burrows et al. 2005). 

Both GRB050406 (Cummings et al. 2005) and GRB050607 (Krimm et al.
2005) show similar behaviour to GRB050502b, in that they have a
re-brightening of their X-ray emission a few hundred seconds after the
initial burst. In both cases, however, the X-ray flares are much
weaker both relative to their respective bursts and to the bright
flare seen in GRB050502b. All three of these bursts are relatively low
in GRB fluence, particularly GRB050406, but are not unusually
faint. Here we report an analysis of all three GRBs, but concentrate
on GRB050502b. As we do not know the redshifts of these bursts, all
physical parameters are quoted in the observer's frame of reference.

The spectrum of the X-ray flare in GRB050502b can be well fitted by a
power law with $\Gamma=2.2\pm 0.03$ but with a time-dependent
absorbing column. Similar evidence for spectral variability is seen
during the flares for the other GRBs, but cannot be well constrained.
In \grb \ the column is highest at the start of the flare, (${\rm N_H}
\approx 1\times 10^{21}$ cm$^{-2}$ above Galactic)
and then decreases strongly suggesting ionization (Burrows et al.
2005). This material was not ionized by the GRB so presumably
is off the line of sight illuminated by the initial jet.
The absorbing column was derived assuming zero redshift so the
intrinsic column will be higher. To derive the X-ray fluence we
performed time-dependent spectral fits for GRB050502b. For the other
two bursts we assumed an intrinsic absorbing column of ${\rm N_H} = 1\times
10^{21}$ cm$^{-2}$ with $\Gamma = 2.4$ and 2.3 for GRB050406 and
GRB050607 respectively. 

In Table 1 we quote fluences in the original detector bandpass and
that derived by extrapolating the power law spectra over the 0.3--350
keV bandpass (all bandpasses quoted are in the observer's frame).
Extending the energy range to lower energies (e.g. 0.1 keV), would
further enhance the relative strength of the X-ray flare in GRB050502b
(e.g. it would be 30\% brighter if a low energy cutoff of 0.1 keV were
adopted).

Our spectral analysis of the X-ray flare in GRB050502b shows that the
total fluence of the flare is comparable to or higher than that of the
initial GRB (Table 1). This is also illustrated in Figure 1 where we
show the flux light curve derived for the XRT bandpass (0.3--10 keV).
For the other GRBs the fluence of the X-ray flare is $\le 10$\% of the initial
burst.

\begin{deluxetable}{lccccl}
\tablecaption{GRB fluences for the initial GRB and the X-ray flare}
\tablecolumns{6}
\tablehead{
\colhead{GRB} & 
\colhead{Burst} & \colhead{Flare} &
\colhead{Burst} & \colhead{Flare} & \colhead{References} \\
\colhead{} & \colhead{($10^{-7}$ erg cm$^{-2}$)} &
\colhead{($10^{-7}$ erg cm$^{-2}$)} &
\colhead{($10^{-7}$ erg cm$^{-2}$)} &
\colhead{($10^{-7}$ erg cm$^{-2}$)} & \colhead{}\\
\colhead{} & \colhead{15--350 keV} & \colhead{0.3--10keV} &
\colhead{0.3--350 keV} & \colhead{0.3--350keV} & \colhead{}
}
\startdata
050502b & 8.0 & 8.8 & 11 & 14 & 1, 2\\
050406  & 9.0 & 0.9 & 57 & 1.1  & 1, 3 \\
050607  & 8.9 & 1.1 & 17 & 1.5 & 1, 4\\
\enddata
\tablerefs{References --- (1) This paper; (2) Cummings et al. 2005; 
(3) Krimm et al. 2005; (4) Retter et al. 2005}
\end{deluxetable}

\begin{figure}
\centerline{\includegraphics[scale=0.6,angle=270]{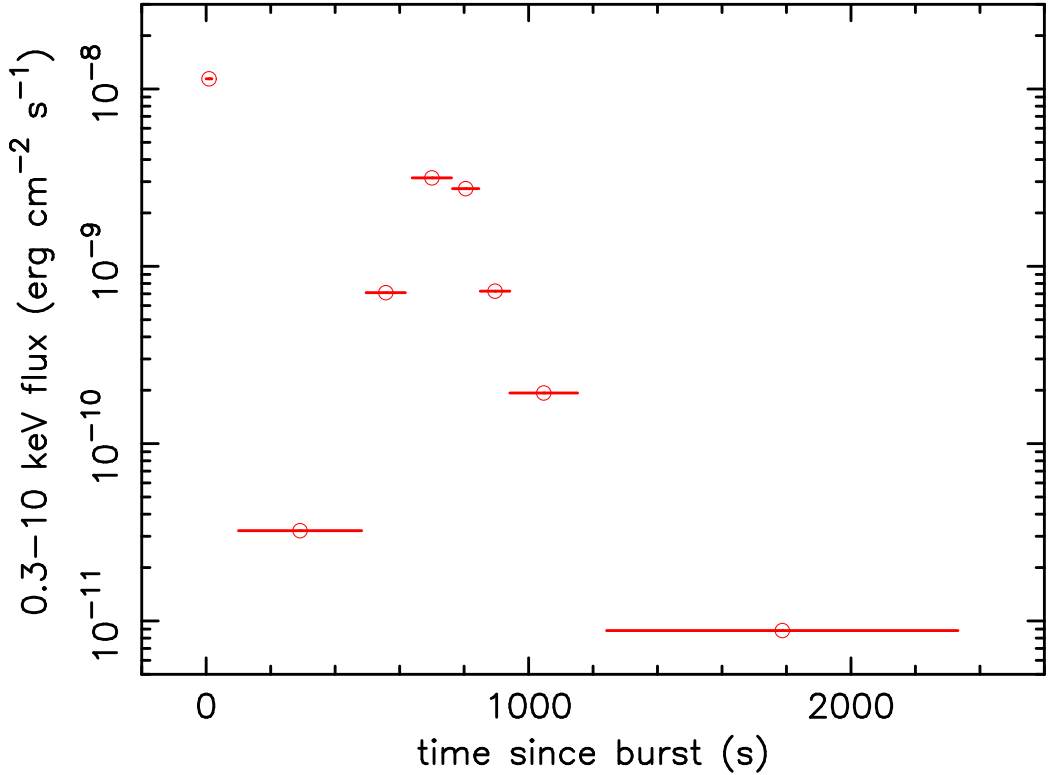}}
\caption{
The unabsorbed 0.3-10 keV flux light curve for the early phase of
GRB050502b. The horizontal bars represent the time intervals over
which the fluxes were calculated. The first point is the initial GRB
extrapolated to the 0.3-10 keV band.}
\end{figure}

\section{Energetics}

We now consider how the central engine in a GRB might restart. 
Our line of argument is straightforward. The original GRB in 050502b
signalled the formation of a compact object (neutron star or black
hole) and the accretion of a stellar mass on to it on a very short
timescale, presumably from a neutron torus. The simplest explanation
for the comparable energy of the X-ray flare is that a second, similar
mass accreted after a delay $t_d$ of up to 400~s (allowing for
possible light-travel effects). Clearly this large mass cannot have
already been in the first torus or it would have been accreted along
with it. Thus a second stellar-mass torus must form and accrete after
$t_d$. 

In 2D models of fallback the extra mass is already in a torus, by
construction, and the delay between the GRB and X--ray flare in \grb \
would have to reflect some viscous angular momentum transport
timescale before the torus became dynamically unstable and rapidly
accreted. In contrast if we do not assume high symmetry, an obvious
possibility is that this mass is a second compact `star' (i.e. a
self-gravitating neutron lump), formed because the collapsing core had
enough angular momentum to fragment. Gravitational radiation then
drags the lump in towards the first compact object where tides smash
it up into a torus which can be accreted, releasing an energy
comparable to the original GRB. This effectively amounts to
restarting, or re-activating, the central engine at late times.

Re-starting the GRB central engine also accounts for the
extraordinarily rapid rate of increase and decrease in the X-ray flare
which goes as greater than $t^7$ if we use the time ($t$) of the
initial burst as the origin $t=0$. Such rates are hard to accommodate
in the standard fireball model for GRB afterglows and instead strongly
support a re-activated central engine.  The BAT on {\it Swift} did not
detect gamma-rays from the X-ray flare in GRB050502b, possibly because
the flare spectrum was soft and the energy was spread over many
seconds.

%

\section{Core collapse and fragmentation}

The two-stage collapse described above is almost exactly that
envisaged by Davies et al. (2002) to explain why hypernovae are rare
among supernovae, and any possible delays between supernovae and
gamma-ray bursts. Davies et al. (2002) argued by comparison with
theoretical studies of star formation (cf. Bonnell \& Pringle, 1995),
that the collapse of a rapidly rotating stellar core was likely to
lead to fragmentation and the initial formation of more than one
compact object of nuclear density. They considered a case where the
collapses producing these objects make a supernova rather than an
initial gamma-ray burst, i.e. where none of them is surrounded by an
accreting torus. Depending on the way fragmentation occurs, cases
where the initial collapse produces a GRB are clearly possible.

The compact objects subsequently coalesce under the effect of
gravitational radiation. Provided that at least one of them is a
neutron star rather than a black hole, Davies et al. show that there
must be a stage where a neutron torus accretes on to a central object,
thus releasing an energy comparable to the initial burst. Eq. (7) of
that paper shows that gravitational radiation emission drags an
orbiting fragment in, to the point where it disrupts and forms a
neutron torus, on a timescale
\begin{equation}
\tau_{\rm GR} = 0.18\times {(a_0/1000~{\rm km})^4\over
  m_1m_2(m_1+m_2)}~{\rm hours.}
\label{gr}
\end{equation}
Here $a_0$ is the initial circular orbit radius and the central black
hole and orbiting fragment have masses $m_1\msun, m_2\msun$
respectively. This
gives a gravitational radiation delay timescale $\tau_{\rm
GR} = 0.18$~hr = 640 s if the merging masses are $\sim 1\msun$ and
their initial separation is $a_0= 1000$~km, which in turn requires a
specific angular momentum $j = 10^{17}$ cm$^2$ s$^{-1}$ --- exactly
what hypernova models require. This will be true for any required GR
delay (which is also affected by the redshift) since $j \propto
a^{1/2} \propto \tau_{\rm GR}^{1/8}$. One can arrange the
gravitational radiation delay to be shorter than the observed delay if
the X--ray event corresponds to a second faster outflow overtaking the
initial one, although a substantial overtaking delay is rather
contrived.

The initial orbit of the fragment may be somewhat misaligned from the
spin axis of the central black hole. Tidal dissipation in the orbiting
fragment and viscous dissipation in the neutron torus cause rapid
alignment through the Lense--Thirring effect (cf. King et al.,
2005). The second burst (X--ray flare) thus has a jet axis close to
the first one. This may explain why the flare is spectrally softer and
lasts longer than the original gamma--ray burst. In the internal shock
picture, the peak spectral energy scales as $E_p \propto
\Gamma^{-2}t_v^{-1}$ (e.g. Table 2 of Zhang \& Meszaros, 2002), where
$\Gamma$ is the bulk Lorentz factor and $t_v$ the variability
timescale of the outflow. The cleaner environment for the second jet
may reduce its baryon loading and thus increase $\Gamma$. It is also
possible that tidal effects make the second accretion event smoother
than the first (increasing $t_v$), although a full hydrodynamical
calculation is needed to check this.

\section{Smaller X-ray flares}

Several other GRBs have been observed to have significant X-ray flares
superimposed on their fading afterglows (see Table 1 and Piro et al.
2005), although none are nearly as bright or quite as late as seen in
\grb. \grb \ itself has another, smaller X-ray event $\approx 10^5$ s
after the burst (Burrows et al. 2005) with a similar fluence to that
of the event in GRB050607 (some 10\% of the initial GRB). These
smaller flares may be evidence of a similar process involving the
accretion of smaller neutron clumps by the central object: for a
sufficiently complex fragmentation process there may even be several
events. A lower limit to the relative luminosity of such smaller
events to the main GRB arises from the result (Davies et al., 2002)
that the minimum mass of a neutron-rich clump is $\sim 0.2\msun$
(smaller masses make nuclei in their centres). The energies of a
merger event to the original GRB must be in the ratio $r =
\eta_2M_2/\eta_1M_1$, where $M_1, M_2$ are the masses of the first
compact object and the subsequent merging object, and $\eta_1, \eta_2$
are the efficiencies of the collapse and merger events respectively.
For comparable efficiencies a merger must have $r\ga 0.2\msun/M_1 \ga
0.02$ (for $M_1 \la 10\msun$). The observed X-ray/GRB ratio depends on
the spectral index of the emitted flux.

\section{Discussion}

The observed fluence of the late X-ray flare in \grb \ is comparable
to that of the main GRB and is spectrally softer.  Smaller, but
possibly related, flares have also been seen in other bursts. Whether
the unusually bright X-ray flare event in GRB050502b is unique or
simply and extreme example of a pattern of behaviour common to GRBs is
unclear. For example the time of the late X-ray flare in GRB050502b is
within the known range in GRB durations (e.g. Paciesas et al. 1999)
which extends to at least a thousand seconds.  Many bursts, but not
all, show spectral evolution such that they become softer later
(e.g. Norris et al. 1986). It has also been shown that if a GRB engine
is quiescent for a long time the subsequent emission outburst could be
unusually large (Ramirez-Ruiz and Merloni 2001). The fragmentation
picture we have suggested here may be able to explain these
phenomena. Continuing accretion at a slower rate from a fragment may
be the origin of the long, faint X--ray afterglows seen in many {\it
Swift} GRBs, which probably require the central engine to remain on
for a long time (cf Zhang \& Meszaros 2001). We will investigate this
possibility in a future paper.

We have suggested that X-ray flares may result from the fragmentation
of the collapsing stellar core, and the subsequent merger of a
significant fragment with the most massive compact object formed in
the collapse. This is a departure from the current hypernova picture,
in which only one object is assumed to form directly in the
collapse. However the rapid rotation apparently required to make a
hypernova is known to lead to fragmentation in other situations such
as star formation. A simple test of our idea will come from combining
gamma-ray burst observations with gravitational-wave detections by
LIGO. Long GRBs should be significant sources of gravitational
radiation, with a chirp signal characteristic of a merging binary
system. If this proves successful it would also give a clean
determination of the masses involved, providing major insight into the
process of core collapse in rapidly rotating stars.

\paragraph*{Acknowledgments}

We thank Evert Rol for useful discussions and the referee for a very
helpful report. ARK acknowledges a Royal Society -- Wolfson Research
Merit Award.  Emma Olsson is supported by a European Union Research
and Training Network grant. The {\it Swift} Project and theoretical
astrophysics research at the University of Leicester are supported by
PPARC.

\end{document}